\documentclass{elsart}
 \usepackage{epsfig}
\usepackage{amssymb}

\begin{document}

\begin{frontmatter}

\title{Nonextensive thermodynamics applied to superconductivity}

\author{Lizardo H. C. M. Nunes}\ead{lizardo@if.uff.br}\and\author{E. V. L. de Mello}
\address{ Departamento de F\'{\i}sica, Universidade Federal Fluminense,
          Av. Litor\^anea s/n,
	  Boa Viagem
	  Niter\'oi, Rio de Janeiro,
          24210-340,
          Brazil }

\begin{abstract}
We have shown that the weak-coupling limit superconductors are well described by $ q \sim 1 $, where $ q $ is a real parameter which characterizes the degree of nonextensivity of Tsallis' entropy. Nevertheless, small deviations with respect to $ q = 1 $ provide better agreement when compared with experimental results. We have also shown that the generalized BCS theory with $ q > 1 $ exhibit power-law behavior of several measurable macroscopic functions in the low-temperature regime. These power-law properties are found in many high-$ T_c $ oxides superconductors and motivated us to extend Tsallis’ entropy calculations to these systems. Therefore, we have calculated the phase diagram and the specific heat and we compare our results with the experimental data for the YBa$_2$Cu$_3$O$_{ 7 - \delta } $ family of compounds.
\end{abstract}


\end{frontmatter}


\section{Introduction}
\label{Int}

The superconductivity interaction in high-$ T_c $  oxides remains one of the greatest puz- zle in condensed matter physics, possibly due to their complex stoichiometry. Despite some evidences towards a phonon-mediated interaction, there is still the question of how a single mechanism produces such a rich phenomenology which includes great charge inhomogeneities, possibly in a stripe morphology, that decreases with the doping level, the appearance of a pseudo-gap at a temperature $ T^∗ $ above the superconducting phase transition at $ T_c $~\cite{Timusk1999}, the presence of a remnant antiferromagnetism in the super- conducting regions, and many other exotic properties.
Although the superconductivity mechanism in these materials is not well known, we share the view that some of these properties may be due to a nonextensive effect of the superconducting interaction, as it was the case for the BCS model~\cite{Nunes2001}. In the weak-coupling limit, making use of an approximated Fermi function, we have shown that measurements of the specific heat, ultrasonic attenuation and tunneling experiments for tin (Sn) are better described with $ q = 0.99 $. In this paper we show that some of the properties for the YBa$_2$Cu$_3$O$_{ 7 - \delta } $  family of compounds can be described by a value of $ q > 1 $.

\section{The Method}
\label{Method}

Consider the following Hamiltonian where there is a on-site Coulomb repulsion, $ U $, which stands for the electron correlations, and a phenomenological attractive interaction, which is responsible for the formation of Cooper pairs,
\begin{equation}
    H
    =
    \sum_{ i j \sigma }
    t_{ i j } c^{ \dagger }_{ i \sigma } c_{ j \sigma }
    +
    U \sum_{ i } n_{ i \uparrow } n_{ i \downarrow }
    +
    +
    \sum_{ \langle i j \rangle \sigma \sigma' }
    V_{ i j }
    c^{ \dagger }_{ i \sigma } c^{ \dagger }_{ j \sigma' }
    c_{ j \sigma' } c_{ i \sigma }
    \, .
    \label{H}
\end{equation}
The strength of the interaction shall be treated as an adjustable parameter. Following the BCS method, the self-consistent gap equation is given by
\begin{equation}
    \Delta( { \bf k } ) =
      \sum_{ {\bf k}' }
          \Delta( { \bf k }' ) V_{ k k' }
		    	\left[
			            \frac{ 1 - 2 f_{q}( \beta E( {\bf k }' ) )
                  }{
                  2 E( { \bf k }'  )
                  }
			\right] \, ,
    \label{gap}
\end{equation}
where $ E_{k} = \sqrt{ e^{ 2 }( { \bf k } ) + \Delta^{2}( {\bf k } ) } $
and $ f_{ q } $ is the generalized Fermi function.
This gap energy may
be considered as the order parameter of the phase transition.
According to experimental data available order parameter is anisotropic in agreement with $ d $-wave superconductor,
$ \Delta( {\bf k } )
=
\Delta_{o}
[
\cos( k_{x} a_{x} )
-
\cos( k_{y} a_{y})
]
$.
A constant nearest-neighbor attraction which favors the formation of singlet Cooper pairs in the d-wave channel is given by
$
V_{ k k' }
=
V_{o}
[ \cos( k_{x} a_{x} ) - \cos( k_{y} a_{y}) ]
$.
$ a_{x} $ and $ a_{y} $
are lattice parameters given by X-ray experiments.

At $ T = T_c $, the gap equation is zero and
Eq. (\ref{gap})
becomes~\cite{Angilella1996}
\begin{equation}
    1
    +
    V
    \langle
        \cos( k_x a_x ) - \cos( k_y a_y )
    \rangle^2_c
    = 0
     \, ,
     \label{Dk}
\end{equation}
where
$\langle \ \ \rangle_c $
represents the sum over the Brillouin Zone at $ T_c $,
that is, at the temperature of vanishing gap.
Thus, Eq. (\ref{Dk}) furnishes $ T_c $ self-consistently.
On the other hand, the hole concentration is given by the q-expectation value of the number operator in the k state, which is simply written as~\cite{Angilella1996}
\begin{equation}
   \rho
   =
   \langle c^{\dagger}_{k} c_{k} \rangle_{q}
   =
   \frac{1}{2}
   \sum_{ \bf k}
          1 - \frac{ e( {\bf k} )  }{  E( {\bf k} )  }
		      \left\{
          \,
		      1 - 2 f_{q}   \left[\,  \beta E( {\bf k} / 2 )   \,\right]
          \,
		      \right\} \, .
    \label{rho}
\end{equation}
The system formed by Eqs. (\ref{Dk}) and (\ref{rho}) provides the phase diagram.

We use also the simplified version of the $ q $-dependent free fermion distribution function~\cite{Buyukkilic1995}
\begin{equation}
f_q
=
\frac{ 1
}{
\left( \e^{ \beta \gamma_k  }_{ q }\right)^{ - q } + 1
}
\, ,
\end{equation}
where
$
e^x_q
\equiv
\left[\,
1 + \left( 1 - q \right) x
\,\right]^{ 1 / ( 1 -  q )  }
$
is the generalized $ q $-exponential function.
Still, the sums are replaced by integrals
and we use a tight binding dispersion relation for a
2-$ d $ orthorhombic lattice.

To apply this formalism to the experimental data for YBCO,
we will use the dispersion relation given by Arpes experiments
which yields the following values for the hopping parameters~\cite{Schabel1998}:
$ t_1 = 0.28147 $ eV,
$ t_2 / t_1 = 0.492 $,
$ t_3 / t_1 = 0.1575 $ ,
$ t_4 / t_1 = 0.1142 $,
$ t_5 / t_1 = 0.0311 $
and
$ a_x$, $a_y $ are 3.88, 3.82 \AA, respectively.

\section{Comparison with experimental data}
\label{Exp}

\begin{figure}
\centerline{
\includegraphics*
[width=15cm]
{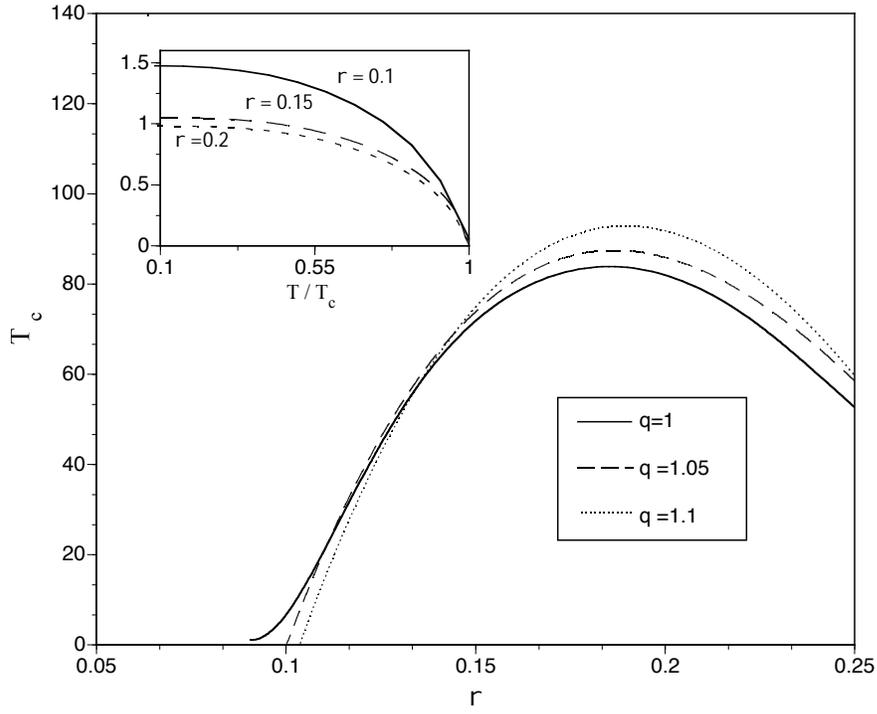}
}
\label{alphaSn}
\caption{Phase diagram of the superconducting phase for different values of $ q $ with $ V = 0.35 t_1 $. Notice that $ T_c $ increases reaching an optimal doping. In the inset, the ratio $ \Delta( T ) = k_B T_c $ is plotted for $ q = 1.05 $. At $ T = 0 $, this ratio varies from $ \approx1.5 $ to $ \approx 1 $.}
\label{Fig1}
\end{figure}

We have solved numerically Eqs. (\ref{Dk}) and (\ref{rho}) self-consistently and then calculated phase diagram is shown in Fig.~\ref{Fig1}.
As the carrier concentration is increased,
$T_c $ increases reaching a maximum at the optimal doping,
for every $ q $,
which is in qualitative agreement with the experimental data~\cite{Tallon1990}.
The critical temperature increases
with $q $ for $ V $ and $ \rho $ fixed,
what means that a smaller $ V $ parameter is required to produce the same effect.
Nevertheless, gap estimations from experimental data available show higher gaps, with $ \Delta(0) / k_B T \sim 2 $,
which is greater than the BCS 1.76 prediction.
The inset shows this ratio varying from
$ \approx 1.5 $
to
$ \approx 1 $
for
$ q = 1.05 $,
for different carrier concentrations.

\begin{figure}
\centerline{
\includegraphics*
[width=15cm]
{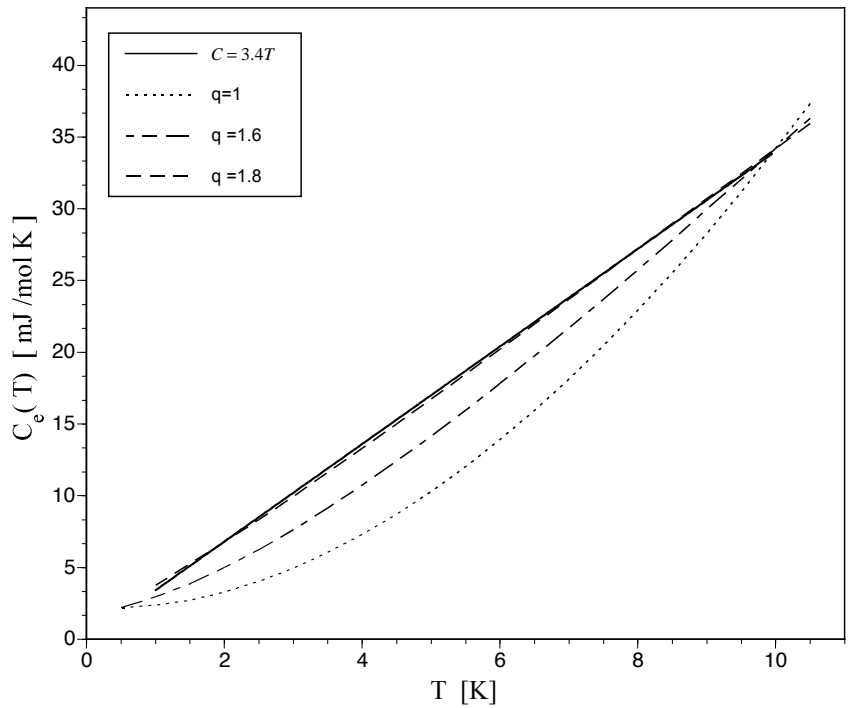}
}
\label{Gapexp}
\caption{Numerical calculations of the low-temperature specific heat adjusted to intercept the fitting proposed in Ref.~\cite{Schabel1998}. As $ q $ increases, they approach the linear-$ T $ behavior found in YBCO samples, with better agreement for $ q = 1.8$.}
\label{Fig2}
\end{figure}

Further, the cuprates present a linear decayment for the electronic specific heat, which is the behavior expected for a free fermion gas.
Experimental data~\cite{Moler1997}
for the twinned YBa$_2$Cu$_2$O$_{ 6.95 } $ sample are fitted by
$ C(T) / T = \gamma +  \beta T^2 $,
where
$ \gamma = 3.4 \, \pm \, 0.03 $ mJ / mol K$ ^2 $,
although most YBCO samples have zero-feld linear-T terms with coefficients
$ \gamma \ge 4 $ mJ / mol K$^2 $,
the other term is related to the lattice contribution.
As shown in Fig.~\ref{Fig2}, our results for
$ q = 1.8 $ have a linear dependence
with temperature in excellent agreement with this fitting,
on the other hand, a quadratic term is what should be expected
for a clean $ d $-wave superconductor;
further, extrinsic contributions,
like those arising from impurities and the Cu–O chains were not taken into account, which indicates that the Tsallis statistics may describe some of the exotic properties found in the high-$T_c $ superconductors.


\newpage

\end{document}